\documentclass[prresearch,twocolumn,superscriptaddress]{revtex4-1}

\usepackage[utf8]{inputenc}
\usepackage[colorlinks, allcolors=blue]{hyperref}
\usepackage{amsmath,revsymb,amssymb}
\usepackage{graphicx}
\usepackage[colorinlistoftodos]{todonotes}
\usepackage{xcolor}
\usepackage{bbold}
\usepackage{braket,ulem}
\usepackage{enumitem}
\usepackage[loose]{units}
\usepackage{color, colortbl}

\DeclareMathOperator{\Tr}{tr}

\renewcommand{\emph}{\textit}

\begin{document}
    
    \title{Experimental demonstration of sequential quantum random access codes}
    
    \author{Giulio Foletto}
    \email{foletto@dei.unipd.it}
    \affiliation{
        Dipartimento di Ingegneria dell'Informazione, Universit\`a di Padova, via Gradenigo 6B, 35131 Padova, Italy}
    \author{Luca Calderaro}
    \affiliation{
        Dipartimento di Ingegneria dell'Informazione, Universit\`a di Padova, via Gradenigo 6B, 35131 Padova, Italy}
    \author{Giuseppe Vallone}
    \affiliation{
    	Dipartimento di Ingegneria dell'Informazione, Universit\`a di Padova, via Gradenigo 6B, 35131 Padova, Italy}
    \affiliation{
    	Dipartimento di Fisica e Astronomia, Universit\`a di Padova, via Marzolo 8, 35131 Padova, Italy}
    \author{Paolo Villoresi}
    \affiliation{
    	Dipartimento di Ingegneria dell'Informazione, Universit\`a di Padova, via Gradenigo 6B, 35131 Padova, Italy}
    
    \begin{abstract}
    	A random access code (RAC) is a strategy to encode a message into a shorter one in a way that any bit of the original can still be recovered with nontrivial probability.
    	Encoding with quantum bits rather than classical ones can improve this probability but has an important limitation: Due to the disturbance caused by standard quantum measurements, qubits cannot be used more than once.
    	However, as recently shown by Mohan, Tavakoli, and Brunner [New J. Phys. {\bf 21} 083034 (2019)], weak measurements can alleviate this problem, allowing two sequential decoders to perform better than with the best classical RAC.
    	We use single photons to experimentally show that these weak measurements are feasible and nonclassical success probabilities are achievable by two decoders.
    	We prove this for different values of the measurement strength and use our experimental results to put tight bounds on them, certifying the accuracy of our setting.
    	This proves the feasibility of using sequential quantum RACs for quantum information tasks such as the self-testing of untrusted devices.
    \end{abstract}
    
    \maketitle
    
    \section{INTRODUCTION}
    A random access code (RAC) is a communication protocol that requires a transmitter (Alice) to encode a $n-$bit long random sequence into a shorter $m-$bit message, and a receiver (Bob) to be able to decode any of the $n$ bits with non-trivial probability $p > 1/2$.
    These parameters are often grouped in expression $n\xrightarrow{p}m$ that describes the task.
    A quantum random access code (QRAC) is the very similar situation in which Alice sends $m$ qubits rather than bits.
    This concept was introduced by Wiesner \cite{Wiesner1983} but caught the interest of the scientific community only after subsequent research by Ambainis \textit{et al.} \cite{Ambainis1999} who showed quantum strategies that achieve $2\xrightarrow{0.85}1$ and $3\xrightarrow{0.78}1$, which beat the best classical RACs for these choices of $n, m$.
    Further studies found that a $4\rightarrow1$ QRAC that reaches $p>1/2$ does not exist \cite{Hayashi2006} but a $4^m-1\rightarrow m$ always does \cite{Iwama}. 
    Other investigations considered different values of $n, m$ \cite{Nayak1999}, the use of qudits ($d-$level quantum systems) rather than qubits \cite{Casaccino2008, Tavakoli2015, Liabotro2017}, or the request of decoding more than 1 bit \cite{Ben-Aroya2008}.
    Applications include communication complexity \cite{Klauck2001}, network coding \cite{Hayashi2007}, locally decodable codes \cite{Kerenidis2004}, dimension witnessing of quantum states \cite{Wehner2008}, self-testing of quantum devices \cite{Tavakoli2018, Farkas2019}, semi-device-independent quantum randomness extraction (SDI-QRE) \cite{Li2012, Lunghi2015, Li2015}, and semi-device-independent key distribution (SDI-QKD) \cite{Pawowski2011, Chaturvedi2018}.
    
    Recently, improvements in the theory and implementation of weak and sequential quantum measurements \cite{Aharonov1988, Mitchison2007, Schiavon2017, Li2018, An2018, Chen2019, Foletto2020}, prompted the introduction of sequential QRACs by Mohan, Tavakoli, and Brunner [(MTB) in what follows] \cite{Mohan2019}.
    Their protocol is a variation of the $2\rightarrow1$ QRAC: Alice encodes a two-bit message into 1 qubit and sends it to Bob, who, after measuring it, forwards the resulting quantum state to a third party (Charlie) who shares the same goal as Bob: decoding any of the two bits of Alice with nontrivial probability $p > 1/2$.
    The core tenets of quantum physics remind us that Bob's measurement disturbs the initial state, making it more difficult for Charlie to extract information from it.
    However, if Bob uses weak measurements rather than projective ones, he can tune this disturbance and give back some information to Charlie at the cost of some of his own.
    This means that Alice's qubit can be used more than once, overcoming a crucial limit of previously studied QRACs, but there is a trade-off between Bob's and Charlie's attainable information that depends on Bob's measurement strength.
    The observation of decoding probabilities that saturate this trade-off self-tests the use of a unique set of states and measurements under the assumption that states are two dimensional and measurements have binary outcomes.
    Additionally, even imperfect results can bind Bob's measurement strength.
    This can be important for the characterization of untrusted quantum devices.
    
    In this paper, we verify MTB's protocol in a quantum optics experiment for different values of the strength parameter.
    We show that it is possible to observe near-optimal decoding probabilities and we put tight bounds on Bob's strength using MTB's self-testing expressions.
    Finally, we discuss some applications of these results.
    \begin{figure*}[t]
    	\centering
    	\includegraphics[width=0.75\textwidth]{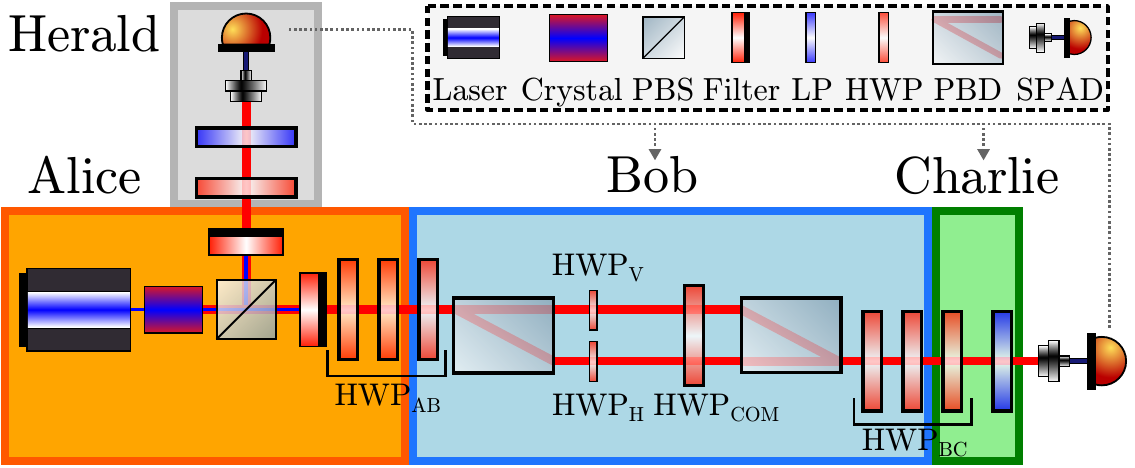}
    	\caption{Scheme of the experimental setup. The sequences of three half-wave plates (HWPs) before and after the Mach-Zehnder interferometer (MZI) are implemented with a single plate each, but we show them here to better separate the roles of Alice, Bob, and Charlie. The arrow indicates that Bob and Charlie only observe the outcome when the detectors click.}
    	\label{fig:ExperimentalScheme}
    \end{figure*} 
    
    \section{MODEL}
    We briefly introduce the quantitative relations presented by MTB and add some comments. 
    Let $x = (x_0, x_1) \in \{0, 1\}^2$ be the two-bit sequence that Alice wants to encode.
    Let $y$ and $z\in\{0, 1\}$ label the positions of the bit in $x$ that Bob and Charlie randomly choose, respectively, to decode.
    Finally, let $b$ and $c$ be the results of Bob and Charlie's respective measurements, associating bit 0 with outcome $+1$ and bit 1 with $-1$.
    We define the two correlation witnesses,
    \begin{align}
    W_{AB} &= \frac18 \sum_{x, y}p(b = x_y|x, y), \label{eq:wab}\\
    W_{AC} &= \frac18 \sum_{x, z}p(c = x_z|x, z), \label{eq:wac}
    \end{align}
    which quantify the probabilities that Bob and Charlie correctly decode the bit they are interested in, averaged over all possible input sequences and bit choices.
    
    If the parties use classical physics, these probabilities are independent of each other and limited by $W_{AB}, W_{AC} \leq \frac34$.
    This upper bound is reached, for example, if Alice sends the first of her bits, meaning that when Bob and Charlie want to decode the second, they can only guess.
    Yet, MTB found that the two decoders can both violate this limit in a quantum scenario. 
    The aforementioned trade-off between the information that each of them can extract translates into an upper bound to $W_{AC}$ that depends on the attained value of $W_{AB}$. 
    In particular,
   	\begin{equation}
    W_{AC} \leq \frac18 \left(4+\sqrt{2} +\sqrt{16W_{AB} -16W_{AB}^2-2} \right),
    \label{eq:tradeoff}
    \end{equation}
    with $W_{AB}$ itself being limited by previous results at $W_{AB}\leq \frac12 +\frac{\sqrt{2}}{4}$ \cite{Ambainis1999}.
    MTB also proposed a strategy to saturate this trade-off and proved that it is unique up to unitary transformations and under the assumption that Alice's state is two dimensional and all measurements have binary outcomes.
    This strategy reads
    \begin{enumerate}[label={C\arabic*}] 
    	\item \label{cond:Alice} Alice encodes her two-bit sequence $x = (x_0, x_1)$ into one of four pure states and sends it to Bob. 
    	These states form the angles of a square in the $XZ$ equatorial line of the Bloch sphere and are equidistant from the eigenstates of $\sigma_X$ and $\sigma_Z$: $\varrho_x = \frac12 \left[\mathbb{1} + (-1)^{x_0}\frac{\sigma_X}{\sqrt{2}} + (-1)^{x_1}\frac{\sigma_Z}{\sqrt{2}} \right]$.
    	\item \label{cond:Bob} Bob weakly measures $\sigma_X$ if $y = 0$ or $\sigma_Z$ if $y = 1$ on the qubit, with strength parameter labeled $\eta \in [0,1]$ as in Ref. \cite{Mohan2019}. The first case ($y=0$) entails using the two-outcome positive operator-valued measure (POVM) $\left(M_{b|0}, b \in \{0,1\} \right)$ where $M_{b|0} = \frac12\left[\mathbb{1} + (-1)^b \eta\sigma_X \right]$. The state is transformed according to Kraus operator $K_{b|0} = \frac12 [ (\cos{\mu} + \sin{\mu} )\mathbb{1} + (-1)^b (\cos{\mu} - \sin{\mu} )\sigma_X  ]$, where $\mu =\frac12 \arccos(\eta)$. In this way, $K_{0|0}^\dagger K_{0|0} - K_{1|0}^\dagger K_{1|0} = M_{0|0} - M_{1|0} = \eta\sigma_X$. The second case ($y=1$) is similar with $\sigma_X$ replaced by $\sigma_Z$.
    	Bob then sends the resulting state to Charlie.
    	\item \label{cond:Charlie} Charlie performs projective measurements of $\sigma_X$ if $z = 0$ or $\sigma_Z$ if $z = 1$.
    \end{enumerate}
	In this situation, the following relations hold:
	\begin{align}
	W_{AB} &= \frac12 + \frac{\sqrt{2}}{4}\eta, \label{eq:wabeta}\\
	W_{AC} &= \frac12 + \frac{\sqrt{2}}{4}\left(\frac{1+\sqrt{1-\eta^2}}{2}\right), \label{eq:waceta}
	\end{align}
	which, when combined, make Expression \eqref{eq:tradeoff} an equality.
	Notably, at least one of these witnesses is always above the classical limit of $\frac34$ and if  $\eta \in \left[\frac{1}{\sqrt2}, \sqrt{2\sqrt2-2} \right]$ both are.
	For $\eta = \frac45$, they take the same value of $\frac12 +\frac{\sqrt{2}}{5}$.
	
	However, we add that this strategy cannot be straightforwardly extended to a third decoder.
	Even if Charlie also uses weak measurements with strength $\eta'$ and relays the resulting qubit to David, there are no values of $(\eta, \eta')$ that provide correlation witnesses greater than $\frac34$ for all three decoders. We show this in Appendix \ref{sec:nomorerecv} finding similar results to those attained in the context of the Clauser-Horne-Shimony-Holt inequality \cite{Mal2016}.
		
	One can wonder whether MTB's protocol can improve the decoding probability of the entire input sequence.
	In a communication scenario in which Bob and Charlie cooperate and agree to always decode different bits, the joint probability of both being correct follows the law:
	\begin{equation}
	\begin{aligned}
	W_{ABC} &= \frac18 \sum_{x, y}p(b = x_y, c = x_z|x, y, z \neq y) \\
	&=\frac14 \left(1+\frac{\eta +\sqrt{1-\eta^2}}{\sqrt{2}} \right).
	\label{eq:success}
	\end{aligned}
	\end{equation}
	It holds that $W_{ABC} \leq \frac12$ with the bound being reached only for $\eta = \frac{1}{\sqrt{2}}$. 
	This agrees with the limits present in the literature: a $m$-qubit system cannot make the decoding probability of a $n$-bit message better than $2^m/2^n$ \cite[Theorem 2.4.2]{Nayak1999}.
	
	The uniqueness of the strategy consisting of \ref{cond:Alice}-\ref{cond:Charlie} allows MTB to conclude that finding $W_{AB}$ and $W_{AC}$ correlated to saturate Expression \eqref{eq:tradeoff} self-tests that the state preparation was that of \ref{cond:Alice} and the measurements were those of \ref{cond:Bob} and \ref{cond:Charlie}. This is an important result for protocols of SDI-QRE or SDI-QKD in which devices cannot be trusted, and their behavior can be checked only from the outcomes they provide.	
	Moreover, even if the values of the witnesses are suboptimal, they still give a lower and an upper bound on parameter $\eta$,
	\begin{align}
	\eta &\geq \eta_{low} = \sqrt{2}\left(2W_{AB}-1\right), \label{eq:etalow}\\
	\eta &\leq \eta_{up} = 2\sqrt{\left(2+\sqrt{2}-4W_{AC}\right)\left(2W_{AC}-1\right)} , \label{eq:etaup}
	\end{align}
	which become tight when conditions \ref{cond:Alice}-\ref{cond:Charlie} are fulfilled.
	These bounds can also be extended to self-tests on the incompatibility between Bob's measurements \cite{Anwer2020}, which is a crucial resource for many quantum information tasks.
	For instance, $W_{AB}$ and $W_{AC}$ can be used as self-tests for the characterization of the QKD state decoders even if the optimal conditions are not reached.

	Finally, we add that trade-off \eqref{eq:tradeoff} and its inverse,
	\begin{equation}
	W_{AB} \leq \frac12 \left[1+\sqrt{4(4+\sqrt{2})W_{AC}-16W_{AC}^2-4-2\sqrt{2}}\right]
	\label{eq:secbound}
	\end{equation} 
	can provide a security bound in an adversarial scenario in which Alice and Charlie try to detect a man in the middle (Bob) or infer the properties of his actions.
	In particular, if $W_{AC} > \frac12 +\frac{\sqrt{2}}{5}\approx 0.783$, then $W_{AC} > W_{AB}$ (see Fig. \ref{fig:witnesses}), meaning that Alice and Charlie can extract a cryptographic key secure from Bob's eavesdropping using a SDI-QKD protocol, such as that of Ref. \cite{Pawowski2011}.
	Compared to the one present in the latter, Eq. \eqref{eq:secbound} is a tighter upper bound on $W_{AB}$ and, in turn, on the mutual information between the legitimate parties' key and the eavesdropper's. Therefore, the performance of the protocol would be increased, although, here, we have the additional assumption that Bob's measurements have binary outcomes.

    \section{METHOD}
    \label{sec:method}
    Our experiment aims at verifying all these relations and showing that it is feasible to meet conditions \ref{cond:Alice}-\ref{cond:Charlie} and find the optimal trade-off.
    We also use Eqs. \eqref{eq:etalow} and \eqref{eq:etaup} to bind the value of $\eta$.
    We choose single photons as our experimental platform and their polarization as the degree of freedom that encodes the information.
    We produce photon pairs at 808 nm through spontaneous parametric down-conversion using a periodically poled potassium titanyl phosphate crystal in a type-II collinear-phase-matching configuration, so that the generated state after the polarizing beam splitter (PBS) is $\ket{\psi}=\ket{H_A}\ket{V_{herald}}$ \cite{Calderaro2018}.
    One photon of each pair is selected in the $\ket{V}$ polarization to filter out imperfections in state preparation and background light, and is sent to a single-photon avalanche diode (SPAD) detector.
    Its presence heralds the other photon of the pair which reaches the core of the setup. 
    
    This is divided into three stages that play the role of Alice, Bob, and Charlie as shown in Fig. \ref{fig:ExperimentalScheme}.
    First, Alice changes the state from $\ket{H}=\Tr_{herald}\left(\ket{\psi}\bra{\psi} \ket{V_{herald}}\bra{V_{herald}}\right)$ to one of the four optimal states of condition \ref{cond:Alice} using a pair of HWPs. 
    Bob carries out the weak measurement with a MZI based on polarizing beam displacers [(PBDs), Thorlabs BD40].
	A first PBD entangles polarization with the path qubit, then the two arms encounter one HWP each, HWP\textsubscript{H} and HWP\textsubscript{V} in Fig. \ref{fig:ExperimentalScheme} with axes at angles 0 and $\pi/4$ relative to the horizontal direction defined by $\ket{H}$.
	HWP\textsubscript{COM} spans across both arms and sets the strength of the measurement through its angle $\theta=\frac{\pi -\arccos(\eta)}{4}$. 
    A second PBD has the dual purpose of closing the interferometer and performing the measurement. It does this by selecting the outcome 0 and sending the corresponding photons to the one exit that continues to the rest of the setup where they meet a HWP at angle $\pi/4$. 
	This MZI + HWP scheme implements $K_{0|0}$: Two more HWPs, one before and one after it, can be rotated to select the other outcome or change the measurement basis.
This means that Bob's apparatus observes one outcome at a time; extensions that allow observing both in separate exits, thus, performing a full measurement, are possible (see Appendix \ref{sec:fullmeasurement}) but beyond the scope of this experiment.
    Charlie's measurements are projective, therefore, his setup consists of a fixed linear polarizer (LP) preceded by a HWP that selects one combination of basis and outcome at a time. 
To reduce the number of components, we replaced the two groups of three consecutive HWPs with a single HWP each, which is controlled by two parties (HWP\textsubscript{AB} and HWP\textsubscript{BC} in Fig. \ref{fig:ExperimentalScheme}).
    Finally, light is coupled into a single-mode fiber and sent to a SPAD detector. Its electrical signals are correlated with those of the herald and coincidences (within a $\pm 1$-ns window) are counted for a fixed exposure time of \unit[2]{s}. 
    The total number of coincidences in this time and for each measurement choice is approximately $8\times 10^3$.

    Our implementation represents a proof of principle demonstration of a QRAC without active random choice of preparation and measurements.
    Moreover, Bob and Charlie do not observe their outcomes independently, but only when the detectors at the end of the setup click.
    We iterate sequentially over all the possible configurations of preparation ($x$), measurement choice ($y, z$), and outcome ($b, c$) by rotating HWP\textsubscript{AB} and HWP\textsubscript{BC}, whose angles are listed in Table \ref{tab:all} (Appendix \ref{sec:expdetails}).
    For each, we record the number of coincident counts.
    These are proportional to the joint probability of the outcomes selected by Bob and Charlie, and we use them to compute the conditional probabilities required by Eqs. \eqref{eq:wab} and \eqref{eq:wac} to find the correlation witnesses.

    \section{RESULTS}
    We measure $W_{AB}$ and $W_{AC}$ for 11 different values of the strength parameter, equally spaced in $[0, 1]$. 
    We use the HWP inside Bob's MZI to set its value of $\eta_{set}$.
    All the results that we report here are extracted from the same experimental data.
    
    \begin{figure}
    	\centering
    	\includegraphics[width=\columnwidth]{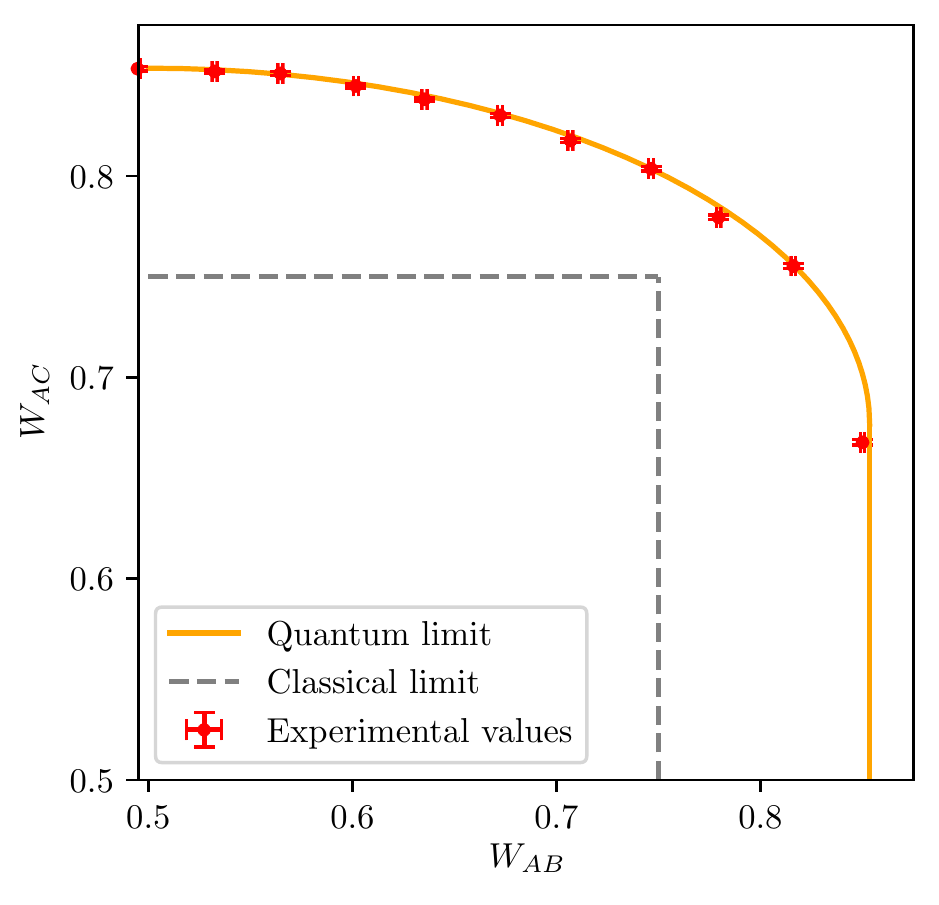}
    	\caption{Experimental correlation witnesses (dots) plotted against each other and compared with the optimal trade-off of Eq. \eqref{eq:tradeoff} (solid line).
    		Here, and in all the following figures, error bars are one standard deviation, obtained from $10^4$ Monte Carlo simulations of the experiment, which consider the Poissonian error on the detected counts.}
    	\label{fig:tradeoff}
    \end{figure} 
    Figure \ref{fig:tradeoff} plots $W_{AC}$ as a function of $W_{AB}$ and compares it with the optimal trade-off that saturates Expression \eqref{eq:tradeoff}.
    The quantum features of the experiment are most evident from the fact that not only all points are outside of the classical region, but also they lie on the boundary of the set of quantum correlations between the witnesses, which certifies that we were able to match the optimal conditions \ref{cond:Alice}-\ref{cond:Charlie}. 
    
    \begin{figure}
    	\centering
    	\includegraphics[width=\columnwidth]{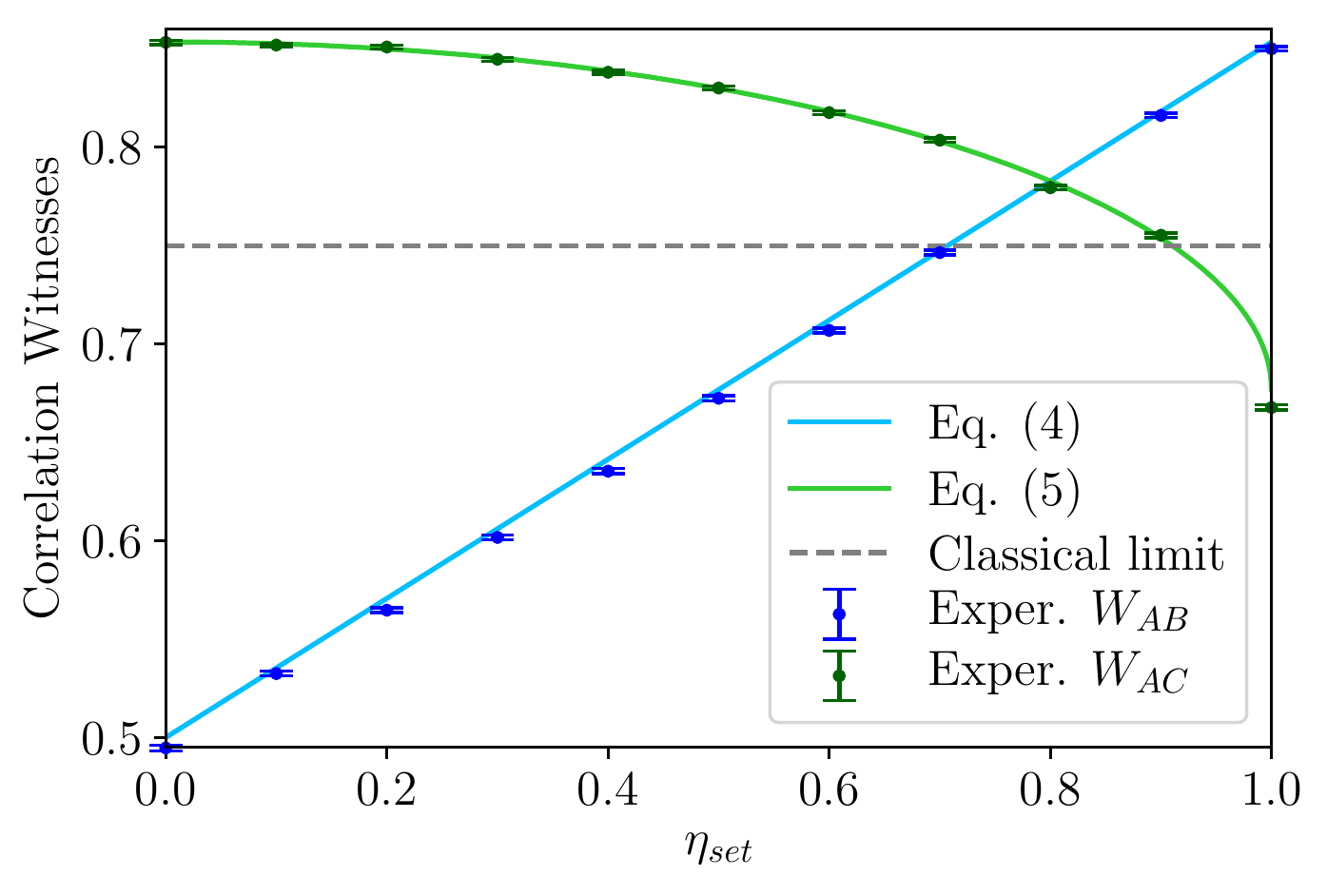}
    	\caption{Experimental correlation witnesses (dots) as a function of the strength parameter that we set using Bob's HWP.
    		We also show the behavior predicted by Eqs. \eqref{eq:wabeta} and \eqref{eq:waceta} (solid lines).
    	 We can see that there is region in which both witnesses are above the classical limit $\left(\frac34\text{, dashed line}\right)$.}
    	\label{fig:witnesses}
    \end{figure} 
	Figure \ref{fig:witnesses} compares the individual witnesses with the expected values of Eqs. \eqref{eq:wabeta} and \eqref{eq:waceta}.
	We clearly see that we could sample the very interesting region in which both $W_{AB}$ and $W_{AC}$ are nonclassical.
	
	\begin{figure}
		\centering
		\includegraphics[width=\columnwidth]{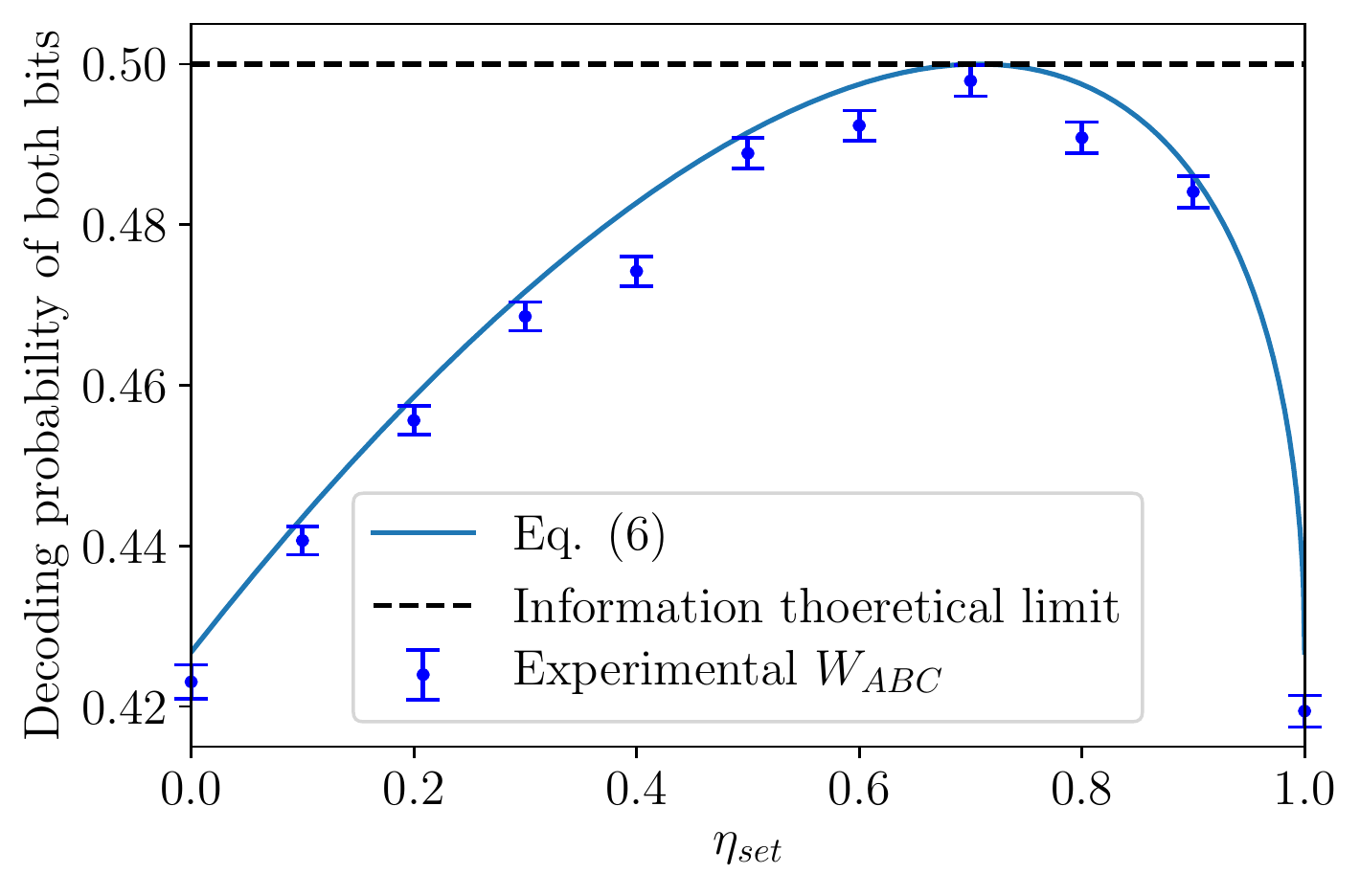}
		\caption{Probability of correctly decoding both of Alice's bits when Bob and Charlie agree to target different bits.}
		\label{fig:success}
	\end{figure} 
	Figure \ref{fig:success} confirms the validity of Eq. \eqref{eq:success} and shows that if Bob and Charlie cooperate to decode the entire input sequence, they cannot succeed with probability better than $\frac12$.
	However, this scheme does allow them to saturate the upper bound for a specific measurement strength.
	
	\begin{figure}
		\centering
		\includegraphics[width=\columnwidth]{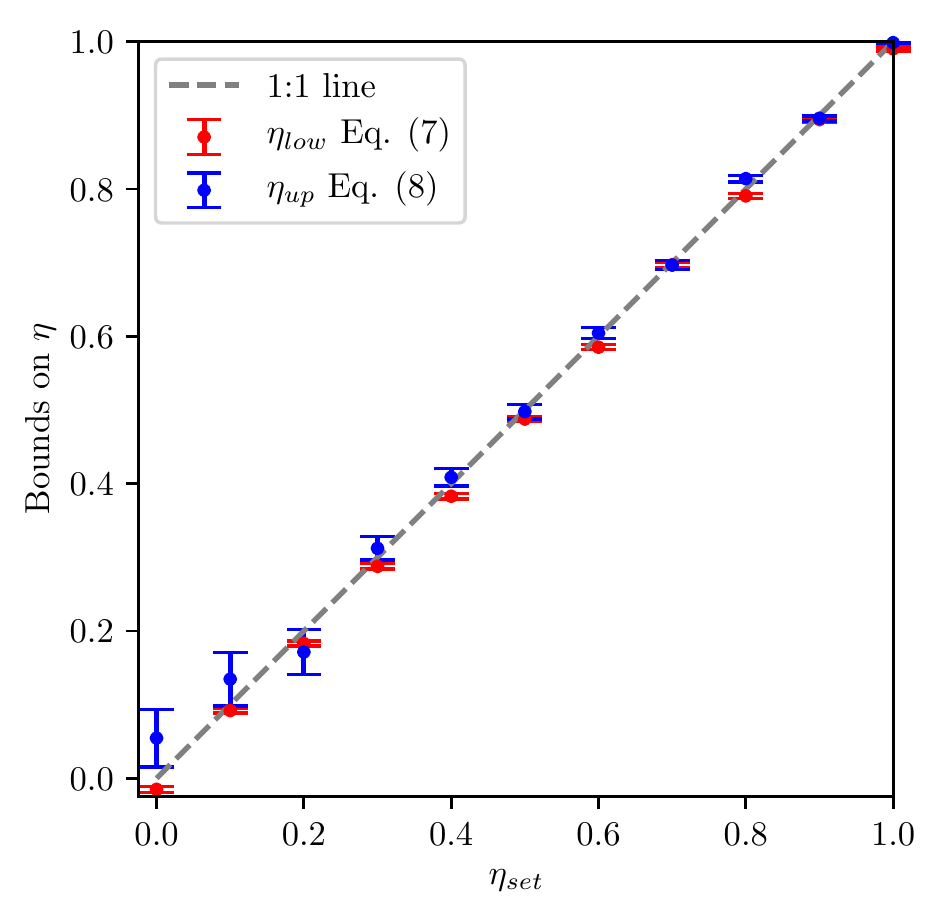}
		\caption{Lower and upper bounds on the strength parameter, obtained by applying relations \eqref{eq:etalow} and \eqref{eq:etaup} to the experimental correlation witnesses.}
		\label{fig:selftest}
	\end{figure} 
	Finally, we evaluated the self-testing capabilities of the protocol, computing upper and lower bounds on $\eta$ from the experimental $W_{AB}$ and $W_{AC}$ using Eqs. \eqref{eq:etalow} and \eqref{eq:etaup}.
	Figure \ref{fig:selftest} plots them as a function of $\eta_{set}$. 
	The tightness of the bounds is another proof that our setup achieved the optimal conditions \ref{cond:Alice}-\ref{cond:Charlie}.
	
    \section{DISCUSSION}
    Our experiment confirms the relations presented by MTB and proves that it is possible for two decoders in a QRAC to share higher success probabilities than admitted by classical physics.
    The quantum weak measurement is the key to this, as it allows reducing the disturbance on the state observed by the first decoder so that it can be used again by the second.
    This is a new situation in which weak measurements prove to be useful and to be able to overcome the limitations of axiomatic projective measurements.
    
    A crucial point of this protocol is that it offers a different way to self-test quantum devices with limited assumptions: Observing the optimal values of $W_{AB}$ and $W_{AC}$ pinpoints (up to unitary transformations) Alice's state preparation and Bob and Charlie's measurements.
    Even without optimality, some properties of Bob's measurements can be bounded.
	This is important for the characterization of setups that implement qubit measurements and require accurate strength setting or exploit incompatibility.
	We have also shown that the concept of sequential QRACs can provide a security bound for a SDI-QKD scenario.
	Additionally, in a communication scenario in which Bob and Charlie cooperate to decode the entirety of Alice's string, there is one value of strength that can reach the performance limit imposed by information theory.
    
    It would also be interesting to study robust self-testing relations for MTB's scheme that can bound other properties of the quantum devices in suboptimal conditions.
    If needed, other assumptions could be added, e.g., perfect knowledge of Alice's preparations could help characterize Bob and Charlie's operations in a measurement-device-independent scenario.
    
    Finally, extensions of Bob's MZI scheme that allow full polarization measurements should be explored, considering also an implementation in integrated optics where polarizing directional couplers and polarization rotators are now feasible and could provide better accuracy than free-space discrete components \cite{Corrielli2014, Gao2015, Sarmiento-Merenguel2015, Pitsios2017}.
   
   \begin{acknowledgments}
   	The authors would like to thank M. Avesani for the useful conversations about information theory and F. Picciariello for his contribution to setting up the experiment.
   	Part of this work was supported by MIUR (Italian Ministry of Education, University and Research) under the initiative ``Departments of Excellence'' (Law No. 232/2016).
   \end{acknowledgments}
    
    \appendix
    \section{EXTENSION OF THE PROTOCOL TO MORE RECEIVERS}
    \label{sec:nomorerecv}
    The protocol can be extended to any number of receivers if they all use weak measurements like Bob.
    However, we show here that this strategy does not allow more than two receivers two achieve correlation witnesses higher than $\frac34$ together.
    Suppose the first receiver (Bob) uses strength parameter $\eta_1$, then,
    \begin{equation}
    \begin{aligned}
    W_{AB}(\eta_1)&= \frac18 \sum_{x,y} \Tr(\varrho_x M_{x_y|y} ) \\
    &= \frac12 + \frac{\sqrt{2}}{4}\eta_1,
    \end{aligned}
    \label{eq:wabfunc}
    \end{equation}
    where $\varrho_x= \frac12 \left[\mathbb{1} + (-1)^{x_0}\frac{\sigma_X}{\sqrt{2}} + (-1)^{x_1}\frac{\sigma_Z}{\sqrt{2}} \right]$ is the state prepared by Alice and $M_{b|y}$ are the operators,
    \begin{equation}
    \begin{aligned}
    M_{0|0} &= \frac12(\mathbb{1} +  \eta_1\sigma_X), \\
    M_{1|0} &= \frac12(\mathbb{1} -  \eta_1\sigma_X), \\
    M_{0|1} &= \frac12(\mathbb{1} +  \eta_1\sigma_Z), \\
    M_{1|1} &= \frac12(\mathbb{1} -  \eta_1\sigma_Z).
    \end{aligned}
    \end{equation}
    Note that $\{M_{0|0}, M_{1|0}\}$ and $\{M_{0|1}, M_{1|1}\}$ are two two-outcome POVMs, indeed, $\sum_b M_{b|y} = \mathbb{1},\ \forall y$.
    Moreover $\sum_b (-1)^b M_{b|0}= \eta_1 \sigma_X$ and $\sum_b (-1)^b M_{b|1}= \eta_1 \sigma_Z$, which is why these POVMs correspond to weak measurements of $\sigma_X$ and $\sigma_Z$, respectively.
    To each $M_{b|y}$ corresponds a Kraus operator $K_{b|y}$ such that $K_{b|y}^\dagger K_{b|y} = M_{b|y}$:
    \begin{equation}
    \begin{aligned}
    K_{0|0} &= \frac12\left[(\cos\mu_1+\sin\mu_1)\mathbb{1} +  (\cos\mu_1+\sin\mu_1)\sigma_X\right], \\
    K_{1|0} &= \frac12\left[(\cos\mu_1+\sin\mu_1)\mathbb{1} -  (\cos\mu_1+\sin\mu_1)\sigma_X\right], \\
    K_{0|1} &= \frac12\left[(\cos\mu_1+\sin\mu_1)\mathbb{1} +  (\cos\mu_1+\sin\mu_1)\sigma_Z\right], \\
    K_{1|1} &= \frac12\left[(\cos\mu_1+\sin\mu_1)\mathbb{1} -  (\cos\mu_1+\sin\mu_1)\sigma_Z\right],
    \end{aligned}
    \end{equation}
    where $\mu_1 = \frac12 \arccos(\eta_1)$.
    
    The second receiver (Charlie) ignores Bob's measurement choice $y$ and outcome $b$, therefore, his correlation witness must be calculated from the postmeasurement state averaged over $y$ and $b$,
    \begin{equation}
    \begin{aligned}
    \varrho_x^B &= \frac12 \sum_{y,b} K_{b|y} \varrho_x K_{b|y}^\dagger \\
    &= \frac12\left[\mathbb{1} +\frac{1+\sqrt{1-\eta_1^2}}{2} \cdot \frac{(-1)^{x_0}\sigma_X + (-1)^{x_1}\sigma_Z}{\sqrt{2}} \right].
    \end{aligned}
    \end{equation}
    This expression is remarkably similar to the initial state $\varrho_x$ but contains factor $\frac{1+\sqrt{1-\eta_1^2}}{2}$ that shortens the Bloch vector of the state. 
    Supposing that Charlie also performs weak measurements with strength parameter $\eta_2$, his correlation witness is:
    \begin{equation}
    \begin{aligned}
    W_{AC}(\eta_1,\eta_2)&= \frac18 \sum_{x,z} \Tr(\varrho_x^B M_{x_z|z} ) \\
    &= \frac12 + \frac{\sqrt{2}}{4}\eta_2\frac{1+\sqrt{1-\eta_1^2}}{2},
    \end{aligned}
    \label{eq:wacfunc}
    \end{equation}
    which coincides with Eq. \eqref{eq:waceta} for $\eta_2 = 1$.
    
    This can continue for any number of receivers and the witness for the $n$th one is as follows:
    \begin{equation}
    W_{AR_n}(\eta_1\ldots\eta_n) = \frac12 + \frac{\sqrt{2}}{4}\eta_n\prod_{i =1}^{n-1} \frac{1+\sqrt{1-\eta_i^2}}{2}.
    \label{eq:warn}
    \end{equation}
    This is an increasing function of $\eta_n$ but a decreasing one of $\eta_i,\ \forall i<n$.
    It can be seen as a generalization of Eq. (15) of Ref. \cite{Mohan2019} and is similar to Eq. (24) of Ref. \cite{Mal2016} (for the case of $n=3$), which was obtained in the context of Bell inequality violations. 
    
    We can see from Eq. \eqref{eq:wabfunc} that $W_{AB}>\frac34$ for $\eta_1=\frac{1}{\sqrt{2}} +\epsilon_1, \ \forall \epsilon_1>0$.
    Plugging this value into Eq. \eqref{eq:wacfunc} shows that $W_{AC}>\frac34$ for $\eta_2=2(\sqrt{2}-1) +\epsilon_2,\ \forall \epsilon_2>(6\sqrt{2}-8)\epsilon_1 +O(\epsilon_1^2)$.
    A third receiver would then find
    \begin{equation}
    \begin{aligned}
    &W_{AR_3}\left[\frac{1}{\sqrt{2}} +\epsilon_1, 2(\sqrt{2}-1) +\epsilon_2, \eta_3\right] \\
    \leq &W_{AR_3}\left[\frac{1}{\sqrt{2}},2(\sqrt{2}-1),1 \right] \\
     =&\frac12+\frac{(\sqrt{2}+1)(1+\sqrt{8\sqrt{2}-11})}{16} \approx 0.735 < \frac34,
    \end{aligned}
    \end{equation}
    where the first inequality is justified by the above monotonicity relations for Eq. \eqref{eq:warn}.
    This means that if Bob and Charlie use measurements strong enough to overcome the classical bound, a third receiver cannot do so, even with maximal strength.
    
    \section{FULL MEASUREMENTS WITH THE MACH-ZEHNDER INTERFEROMETER}
    \label{sec:fullmeasurement}
\begin{figure*}
    	\centering
    	\includegraphics[width=0.8\linewidth]{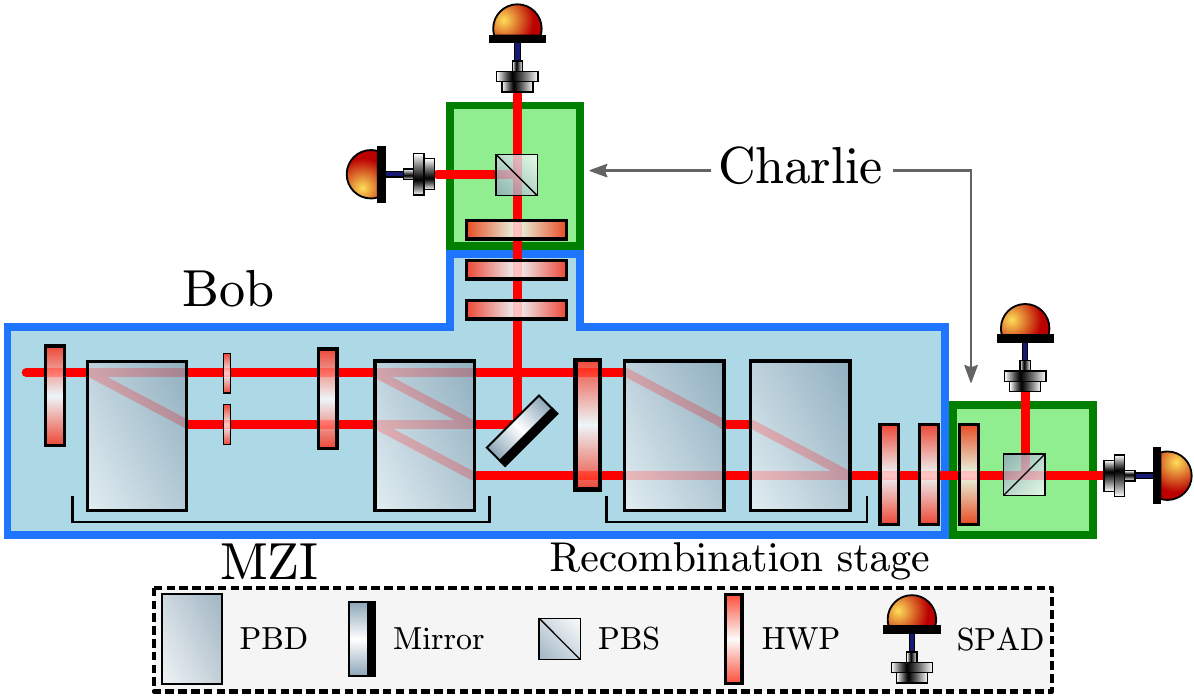}
    	\caption{A possible scheme that extends the MZI used in the experiment to perform a full measurement. Charlie also performs a full measurement by placing detectors at both exits of his PBS.}
    	\label{fig:extension}
    \end{figure*}

    The apparatus made up of a MZI and a HWP that Bob uses to perform the weak polarization measurements can only implement one Kraus operator at a time.
    As described in Sec. \ref{sec:method}, it is possible to switch from one to another by rotating HWPs before and after the MZI.
    However, there are many ways to change the scheme to make a full measurement possible without moving optical components. 
    One is the replacement of PBDs with polarizing beam splitters which would make both exits available.
    A more detailed description of this proposal is in Ref. \cite{Foletto2020}.
    The feasibility of bringing this idea to integrated optics should be explored, because direct translations of PBSs and wave plates exist \cite{Corrielli2014, Pitsios2017} and could allow better accuracy in a much more compact setup.
    However, if implemented with discrete optical table components, this scheme has the disadvantage that PBS-based MZIs are difficult to align.
    A more practical idea is the use of large PBDs that offer three exits, two of which would correspond to the other measurement outcome. They would still need to be recombined with further PBDs, which, if identical to the ones in the MZI, would not ruin the optical coherence.
    The beams could then reach Charlie, who could implement a full measurement using a PBS and two detectors for each input beam.
    HWPs would need to be rotated only to select the measurement basis. Figure \ref{fig:extension} depicts this idea.
    
    Note that, with this scheme, Charlie would know Bob's outcome by observing which detector clicks.
    If this information cannot be simply ignored and must be physically erased, one can imagine to further recombine Bob's exit beams using a (nonpolarizing) beam splitter and delay lines before reaching a single PBS in Charlie's setup.
    
    \section{MORE DETAILS ON THE EXPERIMENT}
    \label{sec:expdetails}
    Table \ref{tab:all} reports the angles of HWP\textsubscript{AB} and HWP\textsubscript{BC} that correspond to each setting of Alice's preparation ($x_0, x_1$), Bob's measurement basis ($y$), Bob's outcome ($b$), Charlie's measurement basis ($z$), and Charlie's outcome ($c$).
    As stated in the main text, HWP\textsubscript{H} and HWP\textsubscript{V} are fixed at angles 0 and $\pi/4$, respectively, whereas HWP\textsubscript{COM} changes only with the measurement strength $\eta$, and its angle is  $\theta=\frac{\pi -\arccos(\eta)}{4}$.
    A description of a very similar setup is also present in Ref. \cite{Foletto2020} with the difference that HWP\textsubscript{H} and HWP\textsubscript{V} are at angles $-\pi/8$ and $\pi/8$, respectively, and $\theta=\pi/8 -\arccos(\eta)/4$.
    Indeed, the MZI works in the same way for any angle $\alpha$ of HWP\textsubscript{V} as long as HWP\textsubscript{H} is at angle $\alpha - \pi/4$ and HWP\textsubscript{COM} is at $\theta=\alpha -\arccos(\eta)/4$.
    The coincident counts observed in the exposure time of \unit[2]{s} for each configuration are also included in Table \ref{tab:all}.
    
    \begin{table*}[]
    	\scriptsize
    	\caption{HWP angles and coincident counts for each configuration of $x_0, x_1, y, b, z, \text{and } c$.}
    	\begin{tabular}{|c|c|c|c|c|c|c|c|c|c|c|c|c|c|c|c|c|c|c|}
    		\hline\hline\rowcolor{white}
    		\multicolumn{6}{|c|}{Settings} & \multicolumn{2}{c|}{Angles (\unit{rad})} & \multicolumn{11}{c|}{Coincident counts in \unit[2]{s}. $\eta_{set} = \ldots$} \\\hline
    		$x_0$	&	$x_1$	&	$y$	&	$b$	&	$z$	&	$c$	&	HWP\textsubscript{AB}	&	HWP\textsubscript{BC}	&	$0$	&	$0.1$	&	$0.2$	&	$0.3$	&	$0.4$	&	$0.5$	&	$0.6$	&	$0.7$	&	$0.8$	&	$0.9$	&	$1$	\\ 
    		\hline\hline\rowcolor{white}
    		0	&	0	&	0	&	0	&	0	&	0	&	$\pi/16$	&	$\pi/4$	&	2864	&	4709	&	5062	&	5313	&	5797	&	5379	&	5779	&	6437	&	6278	&	6517	&	6618	\\\rowcolor[HTML]{E0E0E0}
    		0	&	0	&	0	&	0	&	0	&	1	&	$\pi/16$	&	$\pi/2$	&	450	&	624	&	502	&	461	&	356	&	274	&	219	&	138	&	98	&	36	&	1	\\\rowcolor{white}
    		0	&	0	&	0	&	0	&	1	&	0	&	$\pi/16$	&	$\pi/8$	&	2840	&	4298	&	4461	&	4472	&	4363	&	4277	&	4209	&	4249	&	4137	&	3874	&	3396	\\\rowcolor[HTML]{E0E0E0}
    		0	&	0	&	0	&	0	&	1	&	1	&	$\pi/16$	&	$3\pi/8$	&	529	&	986	&	1200	&	1400	&	1561	&	1638	&	1978	&	2145	&	2483	&	2627	&	3521	\\\rowcolor{white}
    		0	&	0	&	0	&	1	&	0	&	0	&	$5\pi/16$	&	0	&	2988	&	3736	&	3402	&	2885	&	2298	&	1810	&	1486	&	1036	&	685	&	332	&	8	\\\rowcolor[HTML]{E0E0E0}
    		0	&	0	&	0	&	1	&	0	&	1	&	$5\pi/16$	&	$\pi/4$	&	425	&	784	&	792	&	863	&	861	&	801	&	927	&	954	&	1039	&	1021	&	1096	\\\rowcolor{white}
    		0	&	0	&	0	&	1	&	1	&	0	&	$5\pi/16$	&	$-\pi/8$	&	2944	&	4000	&	3798	&	3643	&	3373	&	2925	&	2422	&	2111	&	1709	&	1258	&	511	\\\rowcolor[HTML]{E0E0E0}
    		0	&	0	&	0	&	1	&	1	&	1	&	$5\pi/16$	&	$\pi/8$	&	598	&	657	&	432	&	353	&	177	&	118	&	44	&	16	&	18	&	101	&	563	\\\rowcolor{white}
    		0	&	0	&	1	&	0	&	0	&	0	&	$-\pi/16$	&	$3\pi/8$	&	2852	&	4476	&	4325	&	4556	&	4647	&	3964	&	4233	&	4026	&	4317	&	3797	&	3423	\\\rowcolor[HTML]{E0E0E0}
    		0	&	0	&	1	&	0	&	0	&	1	&	$-\pi/16$	&	$5\pi/8$	&	432	&	846	&	948	&	1172	&	1354	&	1491	&	1714	&	1872	&	2319	&	2459	&	3331	\\\rowcolor{white}
    		0	&	0	&	1	&	0	&	1	&	0	&	$-\pi/16$	&	$\pi/4$	&	2799	&	4492	&	4704	&	5124	&	5454	&	5270	&	5664	&	5809	&	6278	&	6565	&	6454	\\\rowcolor[HTML]{E0E0E0}
    		0	&	0	&	1	&	0	&	1	&	1	&	$-\pi/16$	&	$\pi/2$	&	570	&	716	&	688	&	533	&	434	&	354	&	314	&	209	&	149	&	82	&	1	\\\rowcolor{white}
    		0	&	0	&	1	&	1	&	0	&	0	&	$3\pi/16$	&	$\pi/8$	&	2970	&	4161	&	3992	&	3534	&	3447	&	2407	&	2543	&	2135	&	1931	&	1458	&	632	\\\rowcolor[HTML]{E0E0E0}
    		0	&	0	&	1	&	1	&	0	&	1	&	$3\pi/16$	&	$3\pi/8$	&	475	&	563	&	370	&	241	&	132	&	47	&	21	&	6	&	40	&	124	&	627	\\\rowcolor{white}
    		0	&	0	&	1	&	1	&	1	&	0	&	$3\pi/16$	&	0	&	2967	&	3849	&	3494	&	2881	&	2520	&	1984	&	1657	&	1132	&	816	&	358	&	3	\\\rowcolor[HTML]{E0E0E0}
    		0	&	0	&	1	&	1	&	1	&	1	&	$3\pi/16$	&	$\pi/4$	&	565	&	857	&	907	&	991	&	1035	&	1024	&	1058	&	1087	&	1218	&	1169	&	1227	\\\rowcolor{white}
    		0	&	1	&	0	&	0	&	0	&	0	&	$-\pi/16$	&	$\pi/4$	&	2757	&	4524	&	5024	&	5048	&	5342	&	5312	&	5971	&	6066	&	6399	&	6337	&	6672	\\\rowcolor[HTML]{E0E0E0}
    		0	&	1	&	0	&	0	&	0	&	1	&	$-\pi/16$	&	$\pi/2$	&	554	&	693	&	627	&	548	&	442	&	400	&	293	&	206	&	134	&	71	&	1	\\\rowcolor{white}
    		0	&	1	&	0	&	0	&	1	&	0	&	$-\pi/16$	&	$\pi/8$	&	540	&	886	&	1127	&	1258	&	1451	&	1338	&	1990	&	2062	&	2415	&	2703	&	3462	\\\rowcolor[HTML]{E0E0E0}
    		0	&	1	&	0	&	0	&	1	&	1	&	$-\pi/16$	&	$3\pi/8$	&	3006	&	4363	&	4543	&	4409	&	4451	&	3995	&	4320	&	4021	&	4062	&	3832	&	3522	\\\rowcolor{white}
    		0	&	1	&	0	&	1	&	0	&	0	&	$3\pi/16$	&	0	&	2999	&	3960	&	3516	&	2957	&	2426	&	1949	&	1654	&	1113	&	746	&	403	&	9	\\\rowcolor[HTML]{E0E0E0}
    		0	&	1	&	0	&	1	&	0	&	1	&	$3\pi/16$	&	$\pi/4$	&	561	&	885	&	972	&	962	&	1047	&	939	&	1162	&	1068	&	1205	&	1242	&	1248	\\\rowcolor{white}
    		0	&	1	&	0	&	1	&	1	&	0	&	$3\pi/16$	&	$-\pi/8$	&	456	&	547	&	402	&	281	&	169	&	82	&	23	&	7	&	20	&	129	&	701	\\\rowcolor[HTML]{E0E0E0}
    		0	&	1	&	0	&	1	&	1	&	1	&	$3\pi/16$	&	$\pi/8$	&	2981	&	4118	&	4014	&	3712	&	3373	&	2819	&	2725	&	2168	&	1952	&	1419	&	575	\\\rowcolor{white}
    		0	&	1	&	1	&	0	&	0	&	0	&	$-3\pi/16$	&	$3\pi/8$	&	2889	&	3920	&	3767	&	3441	&	3080	&	2788	&	2441	&	2015	&	1761	&	1300	&	513	\\\rowcolor[HTML]{E0E0E0}
    		0	&	1	&	1	&	0	&	0	&	1	&	$-3\pi/16$	&	$5\pi/8$	&	558	&	677	&	483	&	339	&	198	&	92	&	42	&	10	&	11	&	91	&	617	\\\rowcolor{white}
    		0	&	1	&	1	&	0	&	1	&	0	&	$-3\pi/16$	&	$\pi/4$	&	415	&	692	&	803	&	861	&	899	&	896	&	984	&	980	&	992	&	967	&	1151	\\\rowcolor[HTML]{E0E0E0}
    		0	&	1	&	1	&	0	&	1	&	1	&	$-3\pi/16$	&	$\pi/2$	&	2996	&	3923	&	3428	&	2983	&	2594	&	1868	&	1485	&	1122	&	723	&	339	&	10	\\\rowcolor{white}
    		0	&	1	&	1	&	1	&	0	&	0	&	$\pi/16$	&	$\pi/8$	&	2782	&	4381	&	4514	&	4463	&	4530	&	4358	&	4474	&	4188	&	4445	&	3852	&	3493	\\\rowcolor[HTML]{E0E0E0}
    		0	&	1	&	1	&	1	&	0	&	1	&	$\pi/16$	&	$3\pi/8$	&	575	&	946	&	1152	&	1370	&	1568	&	1664	&	1984	&	2234	&	2556	&	2754	&	3472	\\\rowcolor{white}
    		0	&	1	&	1	&	1	&	1	&	0	&	$\pi/16$	&	0	&	584	&	712	&	625	&	514	&	526	&	390	&	326	&	223	&	168	&	103	&	4	\\\rowcolor[HTML]{E0E0E0}
    		0	&	1	&	1	&	1	&	1	&	1	&	$\pi/16$	&	$\pi/4$	&	2875	&	4625	&	5018	&	5150	&	5543	&	5709	&	6184	&	6094	&	6379	&	6315	&	6570	\\\rowcolor{white}
    		1	&	0	&	0	&	0	&	0	&	0	&	$-5\pi/16$	&	$\pi/4$	&	540	&	854	&	930	&	953	&	1043	&	1008	&	1058	&	1140	&	1170	&	1159	&	1174	\\\rowcolor[HTML]{E0E0E0}
    		1	&	0	&	0	&	0	&	0	&	1	&	$-5\pi/16$	&	$\pi/2$	&	2920	&	3788	&	3365	&	2771	&	2268	&	1794	&	1416	&	1124	&	685	&	343	&	10	\\\rowcolor{white}
    		1	&	0	&	0	&	0	&	1	&	0	&	$-5\pi/16$	&	$\pi/8$	&	2827	&	4005	&	3724	&	3382	&	3210	&	2822	&	2602	&	2215	&	1810	&	1474	&	598	\\\rowcolor[HTML]{E0E0E0}
    		1	&	0	&	0	&	0	&	1	&	1	&	$-5\pi/16$	&	$3\pi/8$	&	443	&	528	&	346	&	242	&	151	&	54	&	16	&	7	&	53	&	147	&	652	\\\rowcolor{white}
    		1	&	0	&	0	&	1	&	0	&	0	&	$-\pi/16$	&	0	&	443	&	580	&	467	&	374	&	352	&	266	&	188	&	150	&	73	&	23	&	6	\\\rowcolor[HTML]{E0E0E0}
    		1	&	0	&	0	&	1	&	0	&	1	&	$-\pi/16$	&	$\pi/4$	&	2757	&	4601	&	4685	&	5179	&	5466	&	5458	&	5724	&	5854	&	6204	&	6432	&	6628	\\\rowcolor{white}
    		1	&	0	&	0	&	1	&	1	&	0	&	$-\pi/16$	&	$-\pi/8$	&	2766	&	4380	&	4347	&	4296	&	4365	&	4190	&	4357	&	4141	&	3959	&	3767	&	3128	\\\rowcolor[HTML]{E0E0E0}
    		1	&	0	&	0	&	1	&	1	&	1	&	$-\pi/16$	&	$\pi/8$	&	376	&	829	&	1007	&	1175	&	1370	&	1542	&	1723	&	1889	&	2290	&	2532	&	3253	\\\rowcolor{white}
    		1	&	0	&	1	&	0	&	0	&	0	&	$-7\pi/16$	&	$3\pi/8$	&	576	&	994	&	1153	&	1290	&	1504	&	1729	&	1860	&	2180	&	2546	&	2725	&	3414	\\\rowcolor[HTML]{E0E0E0}
    		1	&	0	&	1	&	0	&	0	&	1	&	$-7\pi/16$	&	$5\pi/8$	&	2872	&	4271	&	4389	&	4500	&	4339	&	4071	&	4244	&	4275	&	4031	&	4108	&	3212	\\\rowcolor{white}
    		1	&	0	&	1	&	0	&	1	&	0	&	$-7\pi/16$	&	$\pi/4$	&	2979	&	4670	&	4999	&	5214	&	5531	&	5557	&	6258	&	6090	&	6263	&	6660	&	6219	\\\rowcolor[HTML]{E0E0E0}
    		1	&	0	&	1	&	0	&	1	&	1	&	$-7\pi/16$	&	$\pi/2$	&	423	&	571	&	498	&	480	&	392	&	270	&	236	&	152	&	97	&	46	&	5	\\\rowcolor{white}
    		1	&	0	&	1	&	1	&	0	&	0	&	$-3\pi/16$	&	$\pi/8$	&	520	&	551	&	412	&	283	&	150	&	94	&	25	&	1	&	33	&	114	&	586	\\\rowcolor[HTML]{E0E0E0}
    		1	&	0	&	1	&	1	&	0	&	1	&	$-3\pi/16$	&	$3\pi/8$	&	2914	&	4136	&	3730	&	3380	&	2996	&	2664	&	2379	&	1995	&	1757	&	1296	&	523	\\\rowcolor{white}
    		1	&	0	&	1	&	1	&	1	&	0	&	$-3\pi/16$	&	0	&	2997	&	3692	&	3306	&	2869	&	2304	&	1843	&	1542	&	1030	&	669	&	293	&	13	\\\rowcolor[HTML]{E0E0E0}
    		1	&	0	&	1	&	1	&	1	&	1	&	$-3\pi/16$	&	$\pi/4$	&	383	&	686	&	706	&	816	&	816	&	887	&	945	&	961	&	1040	&	1043	&	970	\\\rowcolor{white}
    		1	&	1	&	0	&	0	&	0	&	0	&	$-3\pi/16$	&	$\pi/4$	&	415	&	691	&	744	&	762	&	838	&	860	&	950	&	1013	&	977	&	1059	&	1043	\\\rowcolor[HTML]{E0E0E0}
    		1	&	1	&	0	&	0	&	0	&	1	&	$-3\pi/16$	&	$\pi/2$	&	2975	&	3948	&	3352	&	2930	&	2434	&	1904	&	1535	&	1066	&	755	&	353	&	8	\\\rowcolor{white}
    		1	&	1	&	0	&	0	&	1	&	0	&	$-3\pi/16$	&	$\pi/8$	&	506	&	562	&	406	&	246	&	165	&	92	&	28	&	2	&	30	&	119	&	555	\\\rowcolor[HTML]{E0E0E0}
    		1	&	1	&	0	&	0	&	1	&	1	&	$-3\pi/16$	&	$3\pi/8$	&	2859	&	3977	&	3695	&	3384	&	2984	&	2687	&	2509	&	2011	&	1765	&	1392	&	515	\\\rowcolor{white}
    		1	&	1	&	0	&	1	&	0	&	0	&	$\pi/16$	&	0	&	551	&	776	&	649	&	599	&	430	&	402	&	281	&	268	&	171	&	82	&	6	\\\rowcolor[HTML]{E0E0E0}
    		1	&	1	&	0	&	1	&	0	&	1	&	$\pi/16$	&	$\pi/4$	&	2916	&	4722	&	5020	&	5356	&	5233	&	5407	&	5946	&	6202	&	6538	&	6627	&	6289	\\\rowcolor{white}
    		1	&	1	&	0	&	1	&	1	&	0	&	$\pi/16$	&	$-\pi/8$	&	523	&	935	&	992	&	1230	&	1421	&	1563	&	1901	&	1947	&	2379	&	2728	&	3400	\\\rowcolor[HTML]{E0E0E0}
    		1	&	1	&	0	&	1	&	1	&	1	&	$\pi/16$	&	$\pi/8$	&	2904	&	4386	&	4334	&	4475	&	4119	&	4218	&	4205	&	3970	&	3940	&	3914	&	2964	\\\rowcolor{white}
    		1	&	1	&	1	&	0	&	0	&	0	&	$-5\pi/16$	&	$3\pi/8$	&	462	&	479	&	386	&	228	&	146	&	64	&	15	&	7	&	36	&	119	&	637	\\\rowcolor[HTML]{E0E0E0}
    		1	&	1	&	1	&	0	&	0	&	1	&	$-5\pi/16$	&	$5\pi/8$	&	2942	&	4123	&	3801	&	3568	&	3064	&	2703	&	2678	&	2145	&	1817	&	1458	&	590	\\\rowcolor{white}
    		1	&	1	&	1	&	0	&	1	&	0	&	$-5\pi/16$	&	$\pi/4$	&	542	&	837	&	879	&	986	&	937	&	1034	&	1055	&	1002	&	1139	&	1114	&	1225	\\\rowcolor[HTML]{E0E0E0}
    		1	&	1	&	1	&	0	&	1	&	1	&	$-5\pi/16$	&	$\pi/2$	&	2844	&	3816	&	3398	&	2632	&	2203	&	1872	&	1467	&	1012	&	731	&	317	&	8	\\\rowcolor{white}
    		1	&	1	&	1	&	1	&	0	&	0	&	$-\pi/16$	&	$\pi/8$	&	497	&	910	&	1149	&	1300	&	1466	&	1675	&	1874	&	2055	&	2370	&	2581	&	3239	\\\rowcolor[HTML]{E0E0E0}
    		1	&	1	&	1	&	1	&	0	&	1	&	$-\pi/16$	&	$3\pi/8$	&	2969	&	4372	&	4605	&	4466	&	3927	&	4339	&	4399	&	4043	&	3961	&	3899	&	3235	\\\rowcolor{white}
    		1	&	1	&	1	&	1	&	1	&	0	&	$-\pi/16$	&	0	&	460	&	588	&	466	&	428	&	283	&	238	&	217	&	131	&	77	&	30	&	8	\\\rowcolor[HTML]{E0E0E0}
    		1	&	1	&	1	&	1	&	1	&	1	&	$-\pi/16$	&	$\pi/4$	&	2710	&	4428	&	4619	&	5019	&	4935	&	5534	&	5571	&	5918	&	6186	&	6298	&	6435	\\\rowcolor{white}
    		\hline
    	\end{tabular}
    	\label{tab:all}
    \end{table*}

    %

\end{document}